\documentclass[doublespace]{article}

\usepackage{graphicx,amsmath,lscape,rotating,booktabs}
\usepackage[T1]{fontenc}
\usepackage[utf8]{inputenc}
\usepackage{authblk}
\usepackage{caption}
\usepackage{fullpage}
\usepackage[square,sort]{natbib}
\usepackage{bbm}

\usepackage[colorlinks,bookmarksopen,bookmarksnumbered,citecolor=red,urlcolor=red]{hyperref}
\usepackage{moreverb}
\usepackage{multirow}
\usepackage{rotating}

\usepackage{lscape}
\usepackage[table]{xcolor}
\usepackage{subfig}

\newcolumntype{C}[1]{>{\centering\let\newline\\\arraybackslash\hspace{0pt}}m{#1}}

\begin{document}


\title{ Rethinking non-inferiority: a practical trial design for optimising treatment duration}

\author[1,2]{Matteo Quartagno}
\author[1]{A. Sarah Walker}
\author[1,2]{James R. Carpenter}
\author[1]{Patrick P.J. Phillips}
\author[1]{Mahesh K.B. Parmar}
\affil[1]{MRC Clinical Trials Unit at UCL, Kingsway, London.}
\affil[2]{Department of Medical Statistics, London School of Hygiene
\& Tropical Medicine.}

\date{}
%

\maketitle

\begin{abstract}
\textbf{Background}: trials to identify the minimal effective treatment duration are needed in different therapeutic areas, including bacterial infections, TB and Hepatitis--C. However, standard non-inferiority designs have several limitations, including arbitrariness of non-inferiority margins, choice of research arms and very large sample sizes. 

\textbf{Methods}: we recast the problem of finding an appropriate non-inferior treatment duration in terms of modelling the entire duration-response curve within a pre-specified range. We propose a multi-arm randomised trial design, allocating patients to different treatment durations. We use fractional polynomials and spline-based methods to flexibly model the duration-response curve. We compare different methods in terms of a scaled version of the area between true and estimated prediction curves. We evaluate sensitivity to key design parameters, including sample size, number and position of arms. 

\textbf{Results}: a total sample size of $\sim 500$ patients divided into a moderate number of equidistant arms (5-7) is sufficient to estimate the duration-response curve within a $5\%$ error margin in $95\%$ of the simulations. Fractional polynomials provide similar or better results than spline-based methods in most scenarios.

\textbf{Conclusions}: our proposed practical randomised trial design is an alternative to standard non-inferiority designs, avoiding many of their limitations, and yet being fairly robust to different possible duration-response curves. The trial outcome is the whole duration-response curve, which could be used by clinicians and policy makers to make informed decisions, facilitating a move away from a forced binary hypothesis testing paradigm. 
\end{abstract}

\section{Introduction}\label{sec:intro}

Whilst much early phase drug development is concerned with identifying the most appropriate dose, for many conditions much less emphasis is placed on identifying the most appropriate treatment duration. Consequently, duration is often based as much on precedent as evidence. A motivating example is bacterial infections, where concerns about under-treatment and relatively low costs have historically led to long antibiotic courses. However, the bacteria now causing considerable concern are commensals, carried harmlessly within the normal gut flora, but occasionally causing disease. They can develop resistance when antibiotics are used to treat other conditions (whether required or not, e.g. for viral infections), or for longer than necessary to cure an infection. Widespread antibiotic overuse over the last decades is now considered the main driver for increasing antimicrobial  resistance (AMR)\citep{davies10,ventola15}.

How to design trials to optimise treatment duration (which will often take the
 form of finding the shortest effective treatment duration) is, however, unclear.

The most widely used design is a non-inferiority trial\citep{snapinn00,hahn12}; two key design choices are the new duration of therapy and the non-inferiority margin, i.e. the maximum difference in efficacy of a new versus standard treatment duration, that investigators
will tolerate. If the whole confidence interval (CI) for the difference in
treatment efficacy lies below this margin, non-inferiority of the shorter
duration is demonstrated. However, non-inferiority trials
have been often criticized\citep{rehal16}; key limitations
are:

\begin{itemize}
\item The non-inferiority margin is somewhat arbitrary, typically being a multiple of $5\%$ on the absolute difference scale. European Medicines Agency guidance recommends that the non-inferiority margin for antibiotic trials should be decided so that equivalent efficacy versus placebo can be excluded, e.g. if cure rates are $80\%$ with control and $20\%$ without antibiotics, then the non-inferiority margin should ensure that the intervention has $\geq 20\%$ cure rate. This is rarely helpful, given low cure rates for serious infections without antibiotics, and high cure rates with antibiotics. Further, variation in the typically high cure rates (e.g. between $80-90\%$) can substantially impact the sample size required to demonstrate non-inferiority on an absolute scale. Furthermore, at the design stage, there is relatively little a-priori information on the expected control event rate\citep{head12}; 
\item Whether the CI should be $95\%$ (two-sided alpha=0.05, one-sided alpha=0.025) or $90\%$ (two-sided alpha=0.10, one-sided alpha=0.05) is still debated; 
\item Consequently, sample sizes for non-inferiority trials with reasonably small margins ($5\%$) are usually very large, and often unsuccessful\citep{nimmo15};

\item In our specific exemplar, the shorter duration(s) to be tested have to be chosen in advance; again, limited prior knowledge makes this choice difficult. A bad choice inevitably leads to failure of the trial or an unnecessarily long duration being adopted in clinical practice. Comparing multiple durations increases the chance of selecting sensible durations to test, but requires even bigger sample sizes, with the traditional design;

\item There is no consensus for best analysis methods for non-inferiority trials; both intention-to-treat and per-protocol approach can lead to unreliable results. International recommendations differ\citep{rehal16}; at best, leading to challenges in interpretation and, at worst, to manipulation towards the most favourable results. 

\end{itemize}

An alternative approach to non-inferiority trials is therefore attractive, but relatively little work has been done in this area. 
A recent proposal that has received  considerable attention is the DOOR/RADAR trial design\citep{evans15}.  Response Adjusted for Duration of Antibiotic Risk (RADAR) uses a two-step process: (i) patients are categorized using an overall composite clinical outcome (based on benefits and harms), and (ii) successively ranked with respect to a Desirability Of Outcome Ranking (DOOR), assigning higher ranks to patients with better composite outcomes and shorter durations of antibiotic use. Finally the probability that a randomly selected patient will have a better DOOR if assigned to the new treatment duration is calculated. The main criticisms of DOOR/RADAR are that combining clinical outcome and treatment duration into a single composite outcome may hide important differences in the clinical outcome alone and intrinsically assumes (rather than estimates) that shorter durations are beneficial, and hence the clinical interpretation of the treatment effect on the composite endpoint is far from clear. Phillips et al.\citep{phillips15} showed that two non-inferiority trials where shorter durations had been unequivocally demonstrated \textit{not to be non-inferior} would have instead demonstrated non-inferiority using DOOR/RADAR.

To identify appropriate treatment durations, another possible approach is to model the duration-response curve, borrowing information from other durations when calculating treatment effect at a particular duration. This was first proposed, in a limited way, by Horsburgh et al. \citep{horsburgh13} where, on the log-odds scale, the effect of duration on response rate was assumed to be linear (logistic regression model). 

However, in general, and certainly for antibiotic treatment duration, this strong assumption is unlikely to hold. Therefore, here we instead use flexible regression modelling strategies, specifically fractional polynomials and spline-based methods, to model the duration-response curve, to provide robustness under general forms of the true duration-response curve.

\section{Proposals\label{sec:models}}

Suppose a treatment $T$ has currently recommended duration $D_{max}$ and there is a minimum duration $D_{min}$ we are willing to compare with $D_{max}$, possibly because an even shorter duration is thought unlikely to be sufficiently effective.
Our goal is to model the duration-response curve for response Y between  $D_{min}$ and $D_{max}$. In the equations below, $Y$ can be either a continuous outcome or a linear predictor of a binary outcome (representing cure). In simulations, we will assume $D_{min}=10$ and $D_{max}=20$.    

The most appropriate design depends on the true shape of the duration-response curve; we therefore have to ensure robustness against a series of different scenarios. For example, allocating patients to only two arms, at $D_{max}$ and $D_{min}$ would be a very good design if the duration-response curve was linear, but a terrible design for quadratic duration-response relationships. 

Therefore, instead of focusing on a single duration-response curve, we simulated data from a set of plausible duration-response curves, and then evaluated several study designs across these scenarios. In particular we explored the effect of changing: (i) total sample size $N$, (ii) number and (iii) position of duration arms and (iv) the type of flexible regression model used.

However, in order to select the most accurate procedure for estimating the duration-response curve, we need to choose a measure of discrepancy between the true and estimated curves.

Lack of accuracy is often evaluated through either the integral error or the expected error. For a fixed set of chosen durations $\mathcal{D}=(D_1,\ldots, D_n)=(D_{min},\ldots, D_{max})$, the expected error is defined as:

\begin{equation}
\label{mspe}
EE= \frac{1}{n} \sum_{i=1}^n \Delta(f(D_i),\hat{f}(D_i)) 
\end{equation}

where $\Delta$ represents a sensible measure of distance, e.g. squared difference or absolute difference, $f(D_i)$ represents the true response (typically probability of cure) corresponding to treatment duration $D_i$ and $\hat{f}(D_i)$ represents the corresponding estimate from the fitted model. However, this sum is over the durations defining the support, e.g. only over the specified durations, while we would like to evaluate the fit of the model across the whole duration range $[D_{min},D_{max}]$. Therefore, we instead used a type of integral error, i.e. a measure of accuracy defined though an integral, instead of a sum, to characterize model accuracy over the entire domain of interest $\mathcal{D}=[D_{min},D_{max}]$:

\begin{equation}
\label{integralerror}
IE= \int_{Dmin}^{Dmax} \Delta(f(D),\hat{f}(D)) dD
\end{equation}
 
The D's in this formula are not to be regarded as being sampled randomly from some population, but rather to be considered fixed design variables. We chose the absolute difference as measure of distance $\Delta$, as it has the most straightforward interpretation, namely the area between the true and estimated duration-response curve. Henceforth, we refer to this measure as the ABC (Area Between the Curves). However, ABC has as units probability-days which is challenging to interpret. Therefore, we divided it by $(D_{max}-D_{min})$ to produce a measure on the probability scale, the scaled Area Between the Curves (sABC). For a particular fitted curve, this can be interpreted as a measure of the difference in probability of cure across the whole duration-response curve. Finally, we multiplied this value by 100 and treat it as a percentage in the results we present below. 
 
To model the duration-response curve as flexibly as possible, we compared four different regression strategies:

\begin{enumerate}
\item Fractional Polynomials (FP) \citep{royston94,royston99} of the form:
\begin{equation}
\label{fp}
Y=\beta_1 D^{p_1} + \ldots + \beta_M D^{p_M}.
\end{equation}

with powers $p_1, \ldots, p_M$ taken from a special set $S=\{ -2,-1,-0.5,0,0.5,1,2,3\}$. Usually $M<3$ is sufficient for a good fit; here, we fix $M=2$, producing 36 possible combinations; 
\item Linear splines (LS), with the simplest form, under a single knot K:
\begin{equation}
\label{ls}
Y=\beta_{0}+\beta_{1} D + \beta_2 (D-K)_+ 
\end{equation}
where $(D-K)_+=0$ if $D<K$. 
We investigated linear splines with different numbers of knots; we present results with 3 or 5 knots. Knots are equidistant, within the duration range considered, e.g. for  3 knots (LS3), positioned at $\textbf{K}=\{12.5,15,17.5\}$;
 
\item Linear spline with non-equidistant knots (LSNE): this concentrates knots for the linear splines in the first half of the duration range, where the duration-response relationship is most likely to be non-linear. We use 3 knots, that we arbitrarily chose to position at $\textbf{K}= \{11,13,15\}$;

\item Multivariate Adaptive Regression Splines (MARS) \citep{friedman91,friedman95}, which builds models of the form:

\begin{equation}
\label{MARS}
Y=\sum_{i=1}^k \beta_{i} B_i(D) 
\end{equation}

where each $B_i(D)$ can be (i) a constant, (ii) a hinge function, i.e. $max(0,D-K)$ or $max(0,K-D)$, or (iii) a product of two hinge functions.
A forward selection step, building on a greedy algorithm, is followed by a backward elimination step, to avoid over-fitting. Candidate knots K are all durations observed in the sample, i.e. all selected duration arms.

\end{enumerate}

We did not consider restricted cubic splines\citep{desquilbet10} because preliminary work showed similar results to piece-wise linear splines; therefore we focussed on linear splines for simplicity. Other non-linear regression methods include logistic or Gompertz growth models; however, using these, we would have lost flexibility.

Apart from regression model, other key design parameters are: how many different duration arms should we allocate patients to? How should we space arms across our duration range (e.g. equidistantly)? And, ultimately, how many patients should we enrol? 

To explore these questions, we performed an extensive simulation study, starting from a base-case design and investigating the effect of changing each of the above design parameters in turn.

\section{Results \label{sec:sims}}

The eight different scenarios considered represented a wide range of possible duration-response relationships, from linear to quadratic, sigmoid curves and piecewise functions (Table \ref{tab:scenarios}). We simulated binary responses, representing cure of infection, from a binomial distribution with duration-specific event rates, with 1000 simulated trials for each combination of design parameters.

\begin{table}
\centering
\footnotesize

\begin{tabular}{C{3cm}|C{4.2cm}|C{4cm}|C{5cm}}
Type                                     & Equation                                                   & Characteristics                                                                                             &  Plot
\\
\hline
1. Logistic growth curve  & $P_{success} = 0.05+\dfrac{0.9}{1+\exp(-2D+25)}$             & Increases and asymptotes early&  \raisebox{-1\height}{\includegraphics[width=0.15\textwidth, height=20mm]{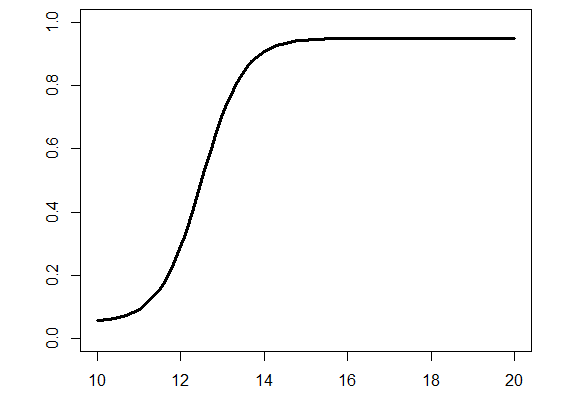}}\\

\hline
2. Gompertz curve A                      & $P_{success} = 0.9\exp(-\exp(-0.5(D-11)))$  &  Small curvature &  \raisebox{-1\height}{\includegraphics[width=0.15\textwidth, height=20mm]{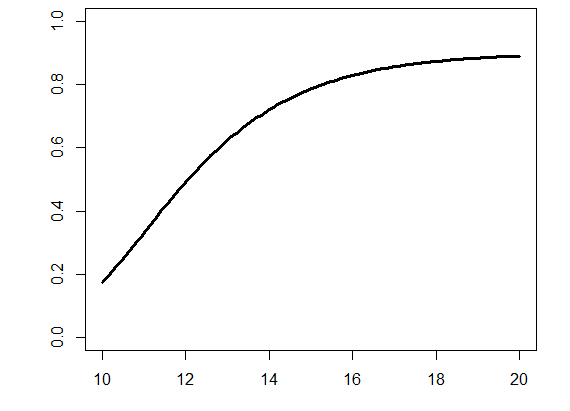}}\\
\hline
3. Gompertz curve B                      & $P_{success} = 0.9\exp(-\exp(-(D-11)))$ & Larger curvature, asymptotes more clearly &  \raisebox{-1\height}{\includegraphics[width=0.15\textwidth, height=20mm]{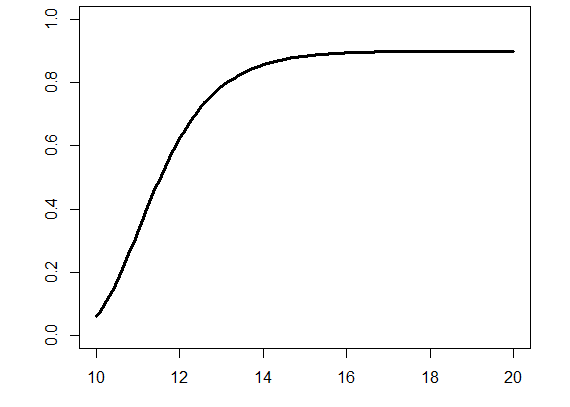}} \\
\hline
4. Gompertz curve C                      & $P_{success} = 0.9\exp(-2\exp(-(D-9)))$ & Asymptotes extremely early &   \raisebox{-1\height}{\includegraphics[width=0.15\textwidth, height=20mm]{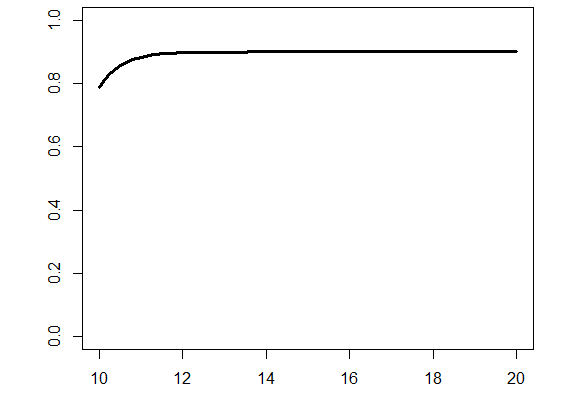}}\\
\hline
5. Linear duration-response curve on logodds scale           & $logit(P_{success})= 0.847+0.210(D-10)$                        & Situation where simple logistic regression is appropriate &  \raisebox{-1\height}{\includegraphics[width=0.15\textwidth, height=20mm]{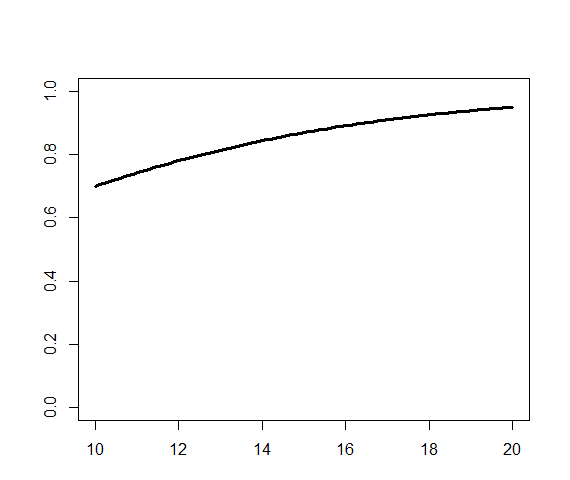}}  \\
\hline
6. Quadratic duration-response curve, curvature\textgreater0 & $P_{success}= 0.7+0.0015(D-10)^2$ & First derivative increasing   & \raisebox{-1\height}{\includegraphics[width=0.15\textwidth, height=20mm]{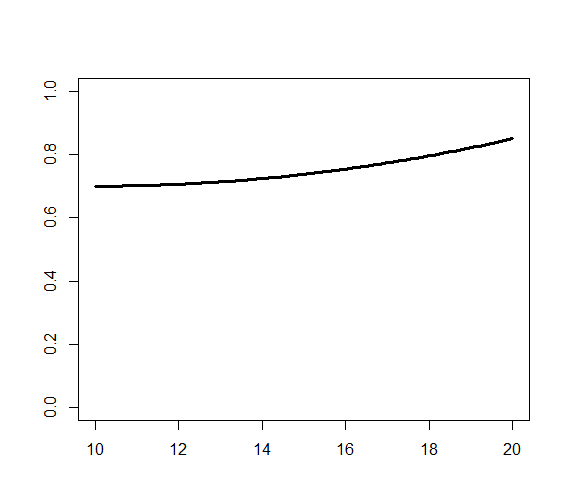}} \\
\hline
7. Quadratic duration-response curve, curvature\textless0    & $P_{success}= 0.7-0.0015(D-10)^2+0.03(D-10)$ & First derivative decreasing  &  \raisebox{-1\height}{\includegraphics[width=0.15\textwidth, height=20mm]{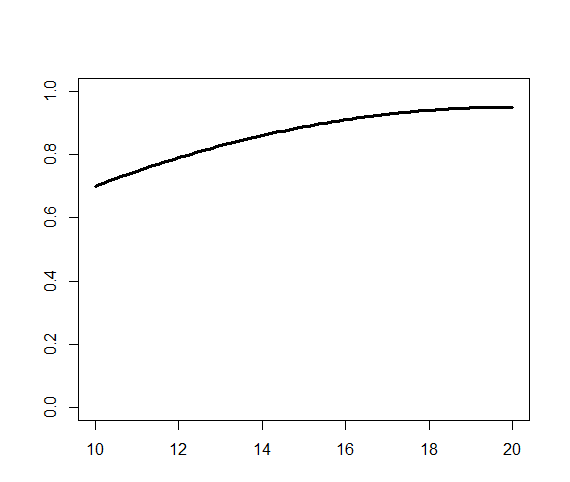}} \\
\hline
8. Piece-wise linear duration-response curve &  $P_{success}$= \newline $(0.5 +0.15(D-10))\mathbbm{1}(D<12) + ( 0.8 +0.05(D-12) ) \mathbbm{1}(D<15) +( 0.94 +0.01(D-15) )\mathbbm{1} (D>15) $     & Different from linear spline logistic regression, here it is linear in the success rate, not in the logodds & \raisebox{-1\height}{\includegraphics[width=0.15\textwidth, height=20mm]{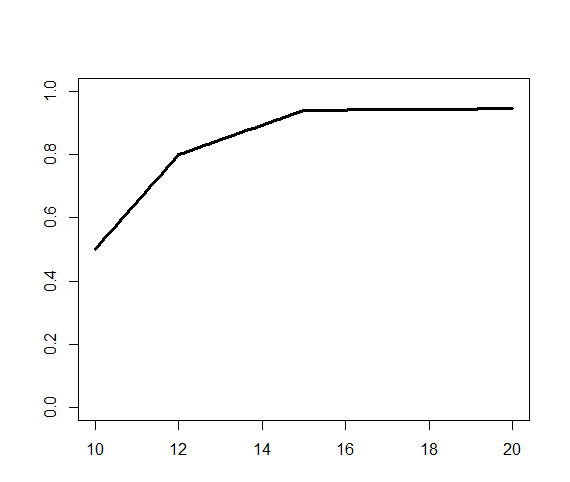}} \\
\hline

\end{tabular}
\caption{Simulation scenarios: eight different data generating mechanisms were investigated. In plots, x axis is treatment duration, and y axis is probability of cure. \label{tab:scenarios}}
\end{table}

\paragraph{Base-case design.}

Here, we fixed a sample size of $504$ individuals randomised  between 7 equidistant duration arms:
$$\textbf{D}=\{10, 11.6, 13.3, 15, 16.6,18.3,20\}$$
Although we could have rounded durations like 11.6 to the closest integer, here we kept them unrounded, simulating a situation where an antibiotic is administered 3 times a day, and therefore $11.6$ means three times daily for eleven days and then twice on the last day. 
Simulated data were analysed with a fractional polynomial logistic regression model, i.e. on the log-odds scale. 

In all 8 scenarios, the worst fit still led to an sABC below $5.3\%$ in $95\%$ of simulations (Table \ref{tab:basecase}), that is, in each scenario $95\%$ of the simulated trials led to an estimated duration-response curve whose error in the estimation of the probability of cure was under $5.3\%$. 

\begin{center}
\begin{figure}
\centering
\includegraphics[width=1\textwidth]{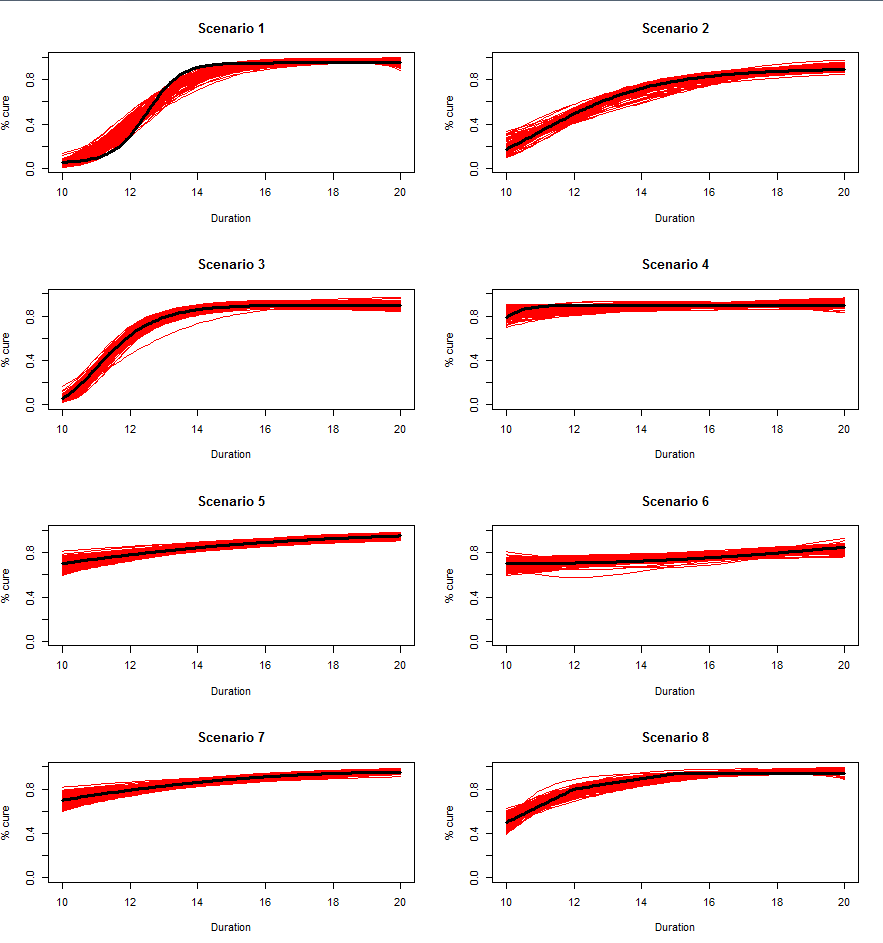}
\caption{Prediction curves (red) from 100 simulations against the true data generating curve (black) for all the eight scenarios. \label{fig:basecase}}
\end{figure}
\end{center}

Scenarios 1, 2 and 3 had the poorest performance. Figure \ref{fig:basecase} shows the fitted prediction curves for a random sample of 100 simulations (red) against the true data generating curve (black). In Scenario 1, fractional polynomials had difficulty in capturing satisfactorily the substantial change in curvature around day 12 and 14, tending to underestimate curvature at these time-points. 

The lowest sABC was obtained with Scenario 5, where the true duration-response curve is linear on the log-odds scale, which is exactly the FP1 model with $p=1$. Similar results were obtained for Scenario 7. 

The maximum sABC was smaller than $10\%$ in all scenarios, meaning that even the simulation leading to the worst fitted prediction curve led to a total bias under $10\%$ in all scenarios.

Finally, visual inspection of Figure \ref{fig:basecase}, seems to confirm that prediction curves look quite similar to the true duration-response curve for a wide variety of scenarios. Next, we investigated the sensitivity of these results to the choice of design parameters and analysis methods.

\begin{table}
\centering

\footnotesize
\hspace*{-1cm}\begin{tabular}{l|ccccc}
           & \multicolumn{5}{c}{sABC}      \\
           & Min & $5^{th}$ percentile & Med. & $95^{th}$ percentile & Max\\
           \hline
Scenario 1 & 0.019 & 0.022 & 0.032 & \textbf{0.051} & 0.077 \\
Scenario 2 & 0.005 & 0.006 & 0.024 & \textbf{0.053} & 0.082 \\
Scenario 3 & 0.003 & 0.007 & 0.022 & \textbf{0.048} & 0.079 \\
Scenario 4 & 0.007 & 0.010 & 0.022 & \textbf{0.039} & 0.050 \\
Scenario 5 & $0.000^*$ & $0.003^*$ & $0.015^*$ & \textbf{0.030}$^*$ & $0.061^*$ \\
Scenario 6 & 0.011 & 0.012 & 0.022 & \textbf{0.044} & 0.066 \\
Scenario 7 & 0.002 & 0.004 & 0.015 & \textbf{0.031} & 0.056 \\
Scenario 8 & 0.009 & 0.010 & 0.025 & \textbf{0.041} & 0.061 \\
\hline
Overall    & 0.000 & 0.006 & 0.022 & \textbf{0.046} & 0.082
\end{tabular}\hspace*{-1cm}
\caption{sABC across the 8 different scenarios in the base-case design (1000 simulations of 504 patients randomised across 7 arms, using FP). Column for the $95^{th}$ sABC is in bold, to show how sABC is smaller, or close to, $5\%$ in all scenarios and overall across all 8000 simulations. Stars next to Scenario 5 results indicate that this is the only scenario where the data generating mechanism is actually a particular case of fractional polynomial on the log-odds scale. \label{tab:basecase}}
\end{table}

\paragraph{Different flexible regression strategies.}

We re-analysed the same simulated data in Table \ref{tab:basecase} using either fractional polynomials (FP), linear spline with 3 or 5 equidistant knots (LS3, LS5), linear spline with knots concentrated in the first half of the curve (LSNE) and multivariate adaptive regression splines (MARS). Only Scenario 5 is the true model for both data generation and analysis. 

\begin{center}
\begin{figure}
\begin{tabular}{cc}
\subfloat[Comparison flexible regression methods: 8000 simulations]{\includegraphics[width=0.5\textwidth]{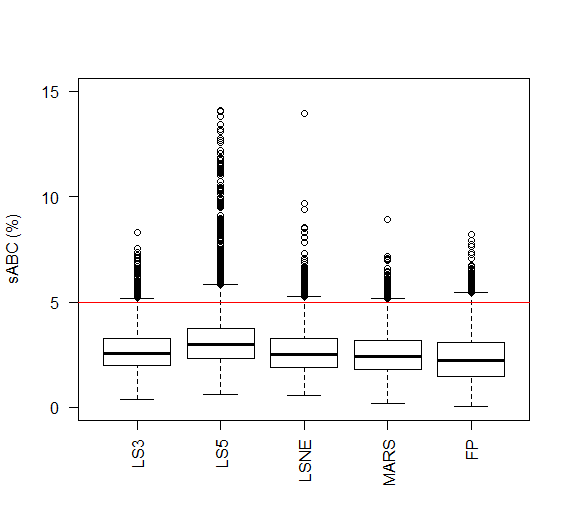}} &
\subfloat[Comparison flexible regression methods: $95^{th}$ percentiles]{\includegraphics[width=0.5\textwidth]{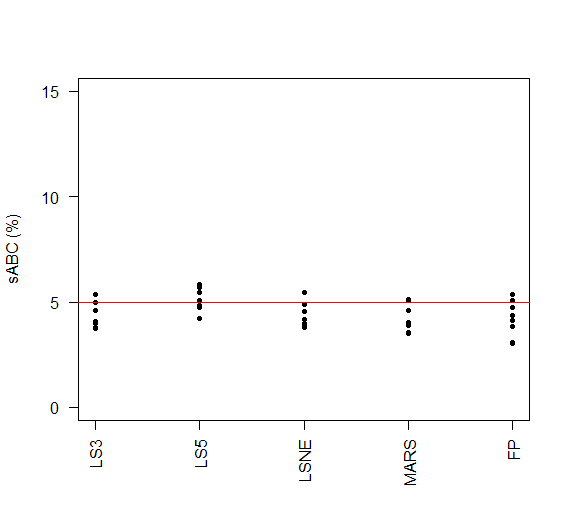}}\\
\subfloat[Sensitivity to sample size: 8000 simulations]{\includegraphics[width=0.5\textwidth]{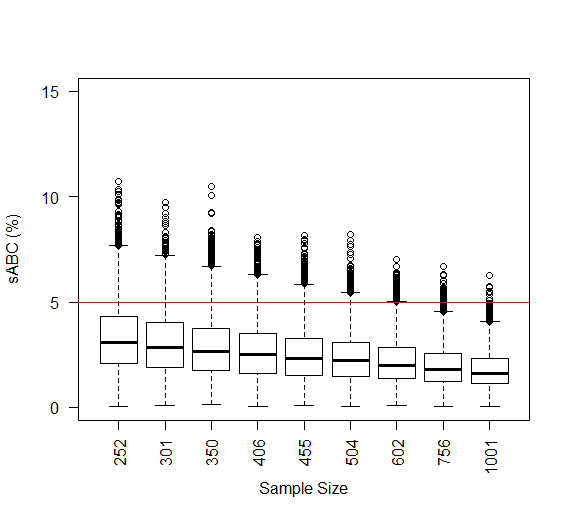}} &
\subfloat[Sensitivity to sample size: $95^{th}$ percentiles]{\includegraphics[width=0.5\textwidth]{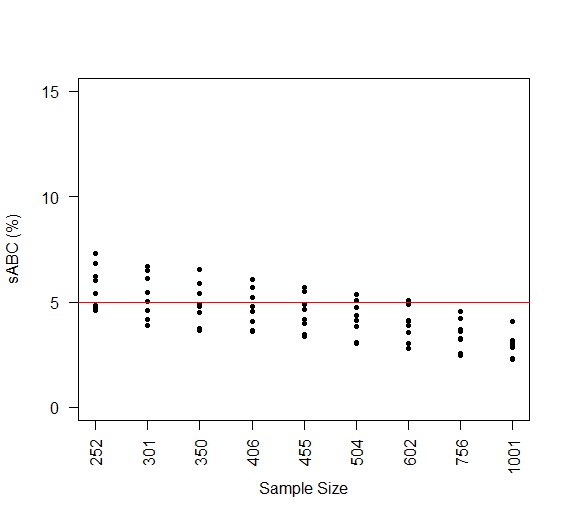}}
\end{tabular}
\caption{Boxplots comparing results of trial simulations from the eight scenarios varying either (i) the flexible regression method used (LS3, LS5, LSNE, MARS, FP), with total sample size of 504 patients (panel \textbf{(a)}), or (ii) the total sample size between 250 and 1000 patients, using FP (panel \textbf{(c)}). Patients are divided in 7 equidistant duration arms. The red horizontal line indicates $5\%$ sABC. In \textbf{(b)} and \textbf{(d)} we compare $95^{th}$ percentiles from the eight scenarios. \label{fig:flexsize}}
\end{figure}
\end{center}

For all methods, sABC for the fitted prediction curves were fairly similar (Figure \ref{fig:flexsize}, (a) and (b)). The only method with slightly inferior performance was LS5, with a slightly larger mean across the 8000 simulations and a higher number of outliers. FP had the smallest mean sABC across the eight scenarios. However, variability between different scenarios was marginally higher than for the other methods.  

\begin{center}
\begin{figure}
\centering
\includegraphics[width=1\textwidth]{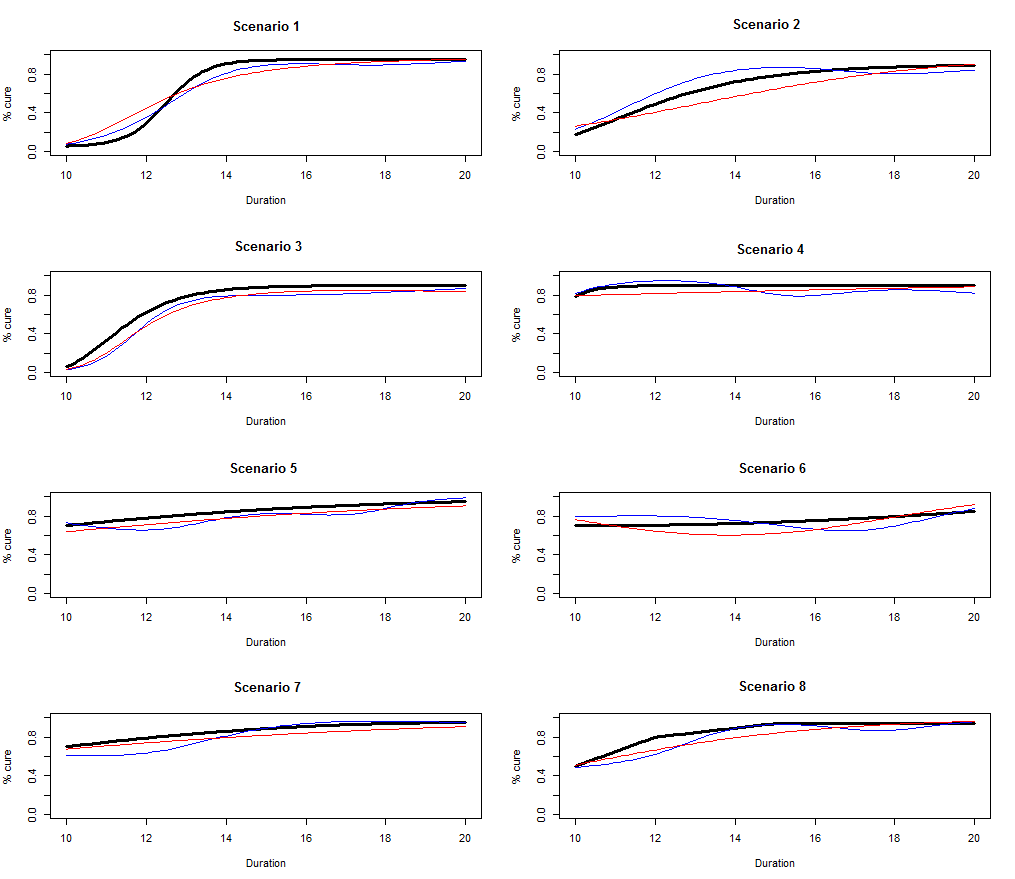}
\caption{ Prediction curves leading to the largest sABC for each of the eight scenarios with the base-case design, analysing data either with LS3 (blue) or FP (red). \label{fig:worst}}
\end{figure}
\end{center}

However FP had an advantage in terms of monotonicity of the curves produced, as shown in Figure \ref{fig:worst}, comparing prediction curves for the simulated dataset with the worst fit (largest sABC), across the eight scenarios, with FP (red) or LS3 (blue). Spline-based methods led to undulating functions, particularly in Scenarios 4,5,6 and 8, while FP prediction curves were smoother and, at least approximately, monotonously increasing, the only exception being the worst fit from Scenario 6.  Spline based methods led to even worse prediction curves in other scenarios, particularly with smaller sample sizes (e.g. 250 patients) and with positions of knots relative to arms chosen in an inconvenient way, e.g. two adjacent knots with no arm in between.  

\paragraph{Total sample size.} 

One  motivation for this study was large sample sizes often required for non-inferiority trials. We therefore investigated the sensitivity of simulation results to total sample sizes varying across $N=(252,301,350,406,455,504,602,756,1001)$, where each number is divisible by 7 (arms).  
 
The second row of Figure \ref{fig:flexsize} summarizes the results in terms of sABC with the different total sample sizes. As expected, increasing total sample size, reduced sABC. With $N=350$ or greater, in more than half the scenarios the $95^{th}$ percentile for sABC was under $5\%$, and in all scenarios for $N\geq 750$. Therefore, above this threshold, whatever the true data-generating curve, in at least $95\%$ of simulated trials we estimated a duration-response curve whose error was lower than $5\%$. 

Figure \ref{fig:flexsize} and Table \ref{tab:basecase} suggest our base-case scenario sample size of 504 might be a reasonable compromise, guaranteeing good estimation of the duration-response curve without requiring too many patients.

\paragraph{Number of duration arms.}

\begin{center}
\begin{figure}
\begin{tabular}{cc}
\subfloat[Sensitivity to number of arms: 8000 simulations]{\includegraphics[width=0.5\textwidth]{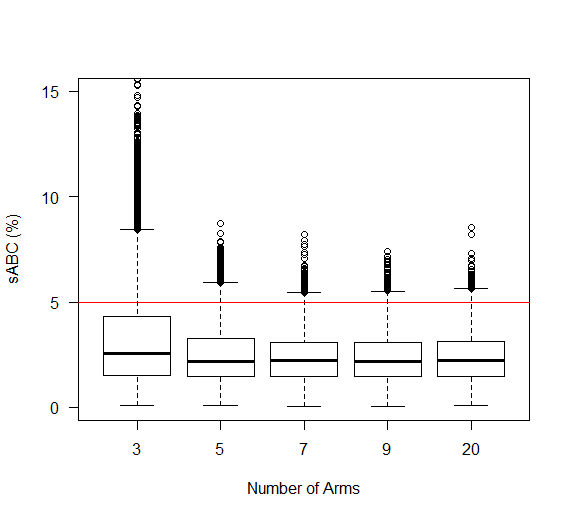}} &
\subfloat[Sensitivity to number of arms: $95^{th}$ percentiles]{\includegraphics[width=0.5\textwidth]{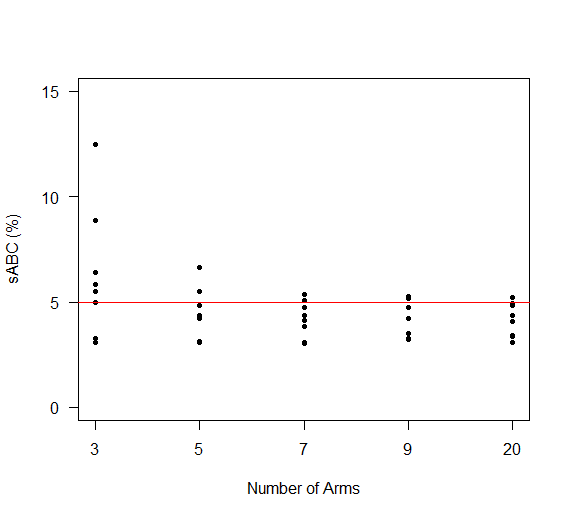}}\\
\subfloat[Sensitivity to placement of arms: 8000 simulations]{\includegraphics[width=0.5\textwidth]{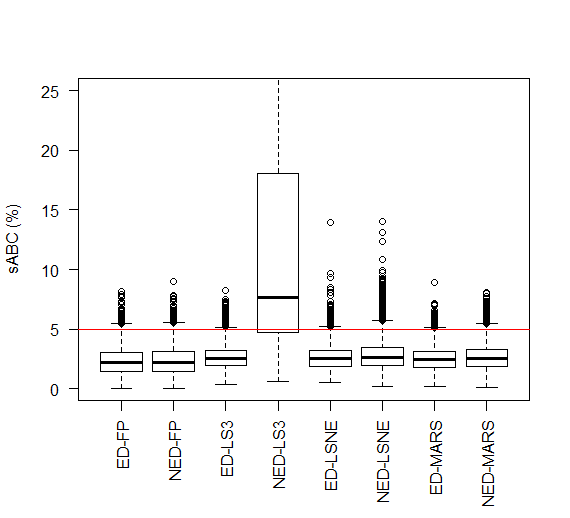}} &
\subfloat[Sensitivity to placement of arms: $95^{th}$ percentiles]{\includegraphics[width=0.5\textwidth]{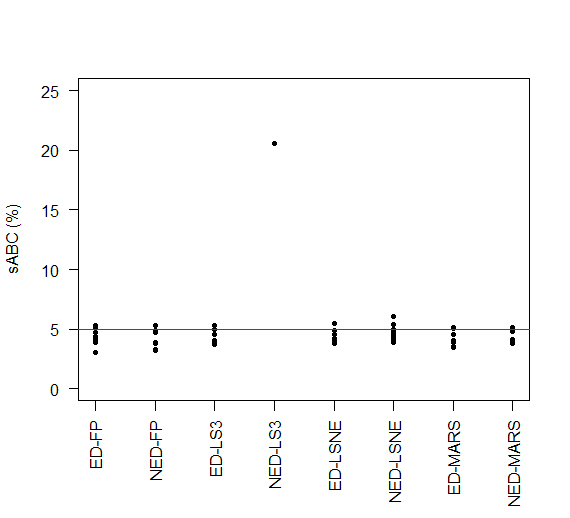}}
\end{tabular}
\caption{Boxplots comparing results of trial simulations from the eight scenarios either varying the number of equidistant arms (panel \textbf{(a)}) between 3 and 20, using FP, or using different designs, ED or NED, comparing four different regression methods (panel \textbf{(c)}). The total sample size is always 504 patients. The red horizontal line indicates $5\%$ sABC. In \textbf{(b)} and \textbf{(d)} we compare $95^{th}$ percentiles from the eight scenarios. In panel (d), there is only one point for NED-LS3, since only in one scenario the $95^{th}$ percentile for sABC was smaller than 0.25. \label{fig:arms}}
\end{figure}
\end{center}

Another obvious question is how many different duration arms should we use. Figures \ref{fig:arms}(a)-(b) compare results from allocating the same number of 500 patients to 3, 5, 9 or 20 arms, rather than the base-case of 7 arms. 

The 3-arm design was clearly inferior and generally led to worse sABC. All other designs led to similar conclusions, and particularly distributions of 7, 9 and 20 arms appeared virtually identical, suggesting that, compared to a base-case of 7 duration arms, we cannot gain much by adding additional arms while keeping sample size fixed.

\paragraph{Position of arms.}

Finally, we investigated the sensitivity of results to the position, rather than the number, of duration arms, by comparing:
\begin{itemize}
\item The standard 7 EquiDistant arms (ED) design;
\item A Not EquiDistant (NED) arms design; this has 5 arms, in the same position as knots of the LSNE spline regression method, i.e. $A=\{10,11,13,15,20\}$. 
\end{itemize}
As above, the motivation for this choice is that the early part of the curve is where the linearity assumption is least likely to hold.

With fractional polynomials, results are similar with the ED and NED design  (Figure \ref{fig:arms} (c)-(d)). This is mainly because the eight scenarios have at most modest departure from linearity in the second half of the curve.

The 3-knot spline regression, performed particularly poorly with the NED design, highlighting an additional issue with spline-based methods versus fractional polynomials, i.e. knot choice. If knots are chosen inappropriately, e.g. two adjacent knots with no arms in between, as for the NED design with the LS3 analysis method, then results may be highly variable. Whilst obvious in this case, similar issues with inappropriate knot positioning might be less trivial to identify in other situations. In contrast, implementation of FP regression is standardised and does not require users to make additional choices.
  
\section{Discussion\label{sec:disc}}

We have proposed a new design for randomised trials to find the effective shorter duration of treatment, for example antibiotics. This is broader development of a previous suggestion\citep{horsburgh13}. The underpinning concept is, instead of directly comparing a limited and arbitrarily chosen number of particular durations, to model the whole duration-response curve across a pre-specified range of durations, in order to maximise the information gained about the effect of shorter or longer regimens. The resulting estimate of the dose-response curve could then be used in a variety of clinically meaningful ways, for example to estimate the minimum duration achieving a certain fixed acceptable cure rate (eg $>90\%$) analogous to a cost-effectiveness acceptability curve (CEAC)\citep{fenwick05}, or combined with other information about toxicity or cost in a decision analytic framework. 
 
Because of lack of information on the true shape of this duration-response curve, we used flexible modelling strategies, in order to protect against misspecification of the parametric form of the model. We compared four different strategies, three based on splines and one on fractional polynomials, concluding that, although spline-based methods can potentially better estimate locally the duration associated with a particular cure rate, fractional polynomials are better at providing a reasonable curve describing the evolution of the cure rate over treatment duration. Fractional polynomials and spline-based methods had been compared in other settings; Binder et al.\citep{binder13} conducted a vast simulation study mimicking biomedical data, broadly concluding that with large datasets the two methods lead to similar results, while in medium-sized datasets FP outperforms spline-based methods on several criteria. They also noted that a major advantage of FP is the simplicity of implementation in standard software packages, compared to the absence of recommendations regarding an appropriate spline based method, matching our conclusions. 

While we could have used FP with more than two polynomials, we focussed on FP2 to reduce the number of parameters, having only a small number of duration arms in our setting. FP and MARS implementation in standard software packages do not allow restriction to monotonously increasing functions; since it is reasonable to assume monotonicity in these trial settings, this might be an area of future research. 

Regarding trial design parameters, a modest number of equidistant arms, e.g. 7, appeared sufficient to give robust results, i.e. the resulting prediction curve from the fit of the model was reasonably close to the true underlying duration-response curve and can therefore provide sufficient information for clinicians about the effect of duration on treatment effect. The NED design provided similar results with only 5 arms (but the same number of patients); however, such a design might be less robust to other shapes of the duration-response curve, e.g. if the curve was far from linearity even in the second part of the duration range investigated. When multi-arm multi-stage designs were first mooted, multiple arms were often raised as a theoretical barrier to recruitment, but subsequent practice has demonstrated that, if anything, these trials are more acceptable to patients, since they ably demonstrate equipoise between a substantial number of treatment options \citep{sydes12}. 

One important issue with flexible regression methods, i.e. particularly in this case  FP and MARS, is accounting for model selection uncertainty \citep{breiman92}. Broadly, the problem is that, since we use the same set of data that we want to analyse to select the final model of interest, the usual standard error estimates from the model tend to be too small. Therefore, a measure of precision of our estimated duration-response curve would require some especially tailored methods, for example bootstrap \citep{buchholz08}. This is a valid theoretical concern: the practical importance of such an adjustment, however, remains to be elucidated.

All designs presented in this study were single stage; however, it would be possible to divide the trial into multiple stages, possibly using some sort of adaptive design, making use of knowledge acquired from previous stages to better design later stages of the trial, for example by dropping some duration arms and/or adding new duration arms. Preliminary simulations investigating this did not suggest substantial gains in accuracy using some simple two-stage designs, but this should be investigated further. 

One legitimate criticism of non-inferiority trials is the arbitrary nature of the non-inferiority margin; in our framework, since $D_{max}$ represents the currently recommended treatment duration, the only arbitrary choice is that of the minimum duration to be considered, $D_{min}$. This choice certainly has a much smaller impact on the trial results than the choice of the non-inferiority margin, but nevertheless it is still extremely important to choose this carefully. Since, as repeated multiple times throughout this paper, we lack any information about the true shape of the duration-response curve below the currently recommended duration, a multi-stage adaptive design could be used to change the position of $D_{min}$ if results after a first stage clearly show this to be too long (i.e. the shortest duration still leading to high efficacy) or too short (i.e. duration extremely ineffective, which might be considered unethical to keep randomising patients to).

Here, we have considered models where the only covariate was treatment duration; however, it would be interesting in the future to expand this methodology by incorporating additional covariate data, such as age and sex. This could be done as a main effect, for example to adjust the minimum duration needed to achieve a threshold cure rate according to other characteristics affecting cure; alternatively, this could be done as an interaction, providing a different duration-response curve for specified subgroups, e.g. males vs females, under 40 vs over 40 years old, etc. Either would allow stratified or personalised medicine, allowing clinicians to prescribe different durations according to key patient characteristics.

The underpinning motivation for this paper was a trial design to identify minimal effective antibiotic treatment duration, but there is no reason why this could not be applied to other similar settings. Examples include trials in TB, where the reason why we might be interested in treatment duration minimization for new regimens is to promote adherence in comparison to a standard of care control, or Hepatitis C where current treatment regimens achieve cure in $>95\%$ of patients but are extremely costly. Alternatively, similar approaches could be applied to dose-intensity of chemotherapy regimens. 

The problem addressed in this paper has similarities with the standard problem in early-phase clinical trials of finding the optimal treatment dose. There is a vast literature on methods for modelling dose-response relationship to find optimal treatment dose \citep{oquigley90, letourneau09}. However, there are important differences making it difficult to use those methods in our situation. The sample sizes required are much smaller in dose-response studies, because the guiding principle is to start with a low dose and to increase it, avoiding exposing too many patients to excessive, and thus unsafe, doses. This is usually done before the drug has actually been tested in phase 2-3 trials. The power of these methods to identify the correct minimum effective dose is therefore often quite low \citep{iasonos08}. In our setting, sample sizes will have to be much larger, because the treatments for which we seek the minimum effective duration are known to be effective and do not have to pass through phase 3 trials afterwards, and therefore we want to be sure not to recommend an insufficiently effective treatment duration. With larger sample sizes, methods like the Continual Reassessment Method (CRM) become infeasible, most of all in the TB example where treatment may last several months. Furthermore, in early-stage trial, the focus is often on pharmacokinetic, and the specific forms of the dose-response curves used usually derive from the underlying pharmacokinetic models for drug absorption into the bloodstream.

In conclusion, we believe that our proposed new paradigm for clinical trials to optimise treatment duration has the potential to revolutionise the design of trials where reducing treatment duration is our goal, e.g. in the fight against AMR. Our approach moves away from multiple inefficient trials of arbitrary antibiotic durations that may all be suboptimal, and avoids the inherent complications of the DOOR/RADAR approach.  We have shown how certain design parameters may affect the fit of a flexible regression strategy to model the duration-response curve. Randomising approximately 500 patients between a moderate number of equidistant arms (5-7) is sufficient under a range of different possible scenarios to give a good fit and describe the duration-response curve well.  
 
\bibliography{biblio}

\begin{thebibliography}{}

\bibitem[Binder et~al., 2013]{binder13}
Binder, H., Sauerbrei, W., and Royston, P. (2013).
\newblock {{C}omparison between splines and fractional polynomials for
  multivariable model building with continuous covariates: a simulation study
  with continuous response}.
\newblock {\em Stat Med}, 32(13):2262--2277.

\bibitem[Breiman, 1992]{breiman92}
Breiman, L. (1992).
\newblock The little bootstrap and other methods for dimensionality selection
  in regression: X-fixed prediction error.
\newblock {\em Journal of the American Statistical Association},
  87(419):738--754.

\bibitem[Buchholz et~al., 2008]{buchholz08}
Buchholz, A., Holländer, N., and Sauerbrei, W. (2008).
\newblock On properties of predictors derived with a two-step bootstrap model
  averaging approach—a simulation study in the linear regression model.
\newblock {\em Computational Statistics \& Data Analysis}, 52(5):2778 -- 2793.

\bibitem[Davies and Davies, 2010]{davies10}
Davies, J. and Davies, D. (2010).
\newblock Origins and evolution of antibiotic resistance.
\newblock {\em Microbiology and Molecular Biology Reviews}, 74(3):417--433.

\bibitem[Desquilbet and Mariotti, 2010]{desquilbet10}
Desquilbet, L. and Mariotti, F. (2010).
\newblock {{D}ose-response analyses using restricted cubic spline functions in
  public health research}.
\newblock {\em Stat Med}, 29(9):1037--1057.

\bibitem[Evans et~al., 2015]{evans15}
Evans, S.~R., Rubin, D., Follmann, D., Pennello, G., Huskins, W.~C., Powers,
  J.~H., Schoenfeld, D., Chuang-Stein, C., Cosgrove, S.~E., Fowler, V.~G.,
  Lautenbach, E., and Chambers, H.~F. (2015).
\newblock {{D}esirability of {O}utcome {R}anking ({D}{O}{O}{R}) and {R}esponse
  {A}djusted for {D}uration of {A}ntibiotic {R}isk ({R}{A}{D}{A}{R})}.
\newblock {\em Clin. Infect. Dis.}, 61(5):800--806.

\bibitem[Fenwick and Byford, 2005]{fenwick05}
Fenwick, E. and Byford, S. (2005).
\newblock A guide to cost-effectiveness acceptability curves.
\newblock {\em The British Journal of Psychiatry}, 187(2):106--108.

\bibitem[Friedman, 1991]{friedman91}
Friedman, J.~H. (1991).
\newblock Multivariate adaptive regression splines.
\newblock {\em The Annals of Statistics}, 19(1):1--67.

\bibitem[Friedman and Roosen, 1995]{friedman95}
Friedman, J.~H. and Roosen, C.~B. (1995).
\newblock {{A}n introduction to multivariate adaptive regression splines}.
\newblock {\em Stat Methods Med Res}, 4(3):197--217.

\bibitem[Hahn, 2012]{hahn12}
Hahn, S. (2012).
\newblock {{U}nderstanding noninferiority trials}.
\newblock {\em Korean J Pediatr}, 55(11):403--407.

\bibitem[Head et~al., 2012]{head12}
Head, S.~J., Kaul, S., Bogers, A.~J., and Kappetein, A.~P. (2012).
\newblock Non-inferiority study design: lessons to be learned from
  cardiovascular trials.
\newblock {\em European Heart Journal}, 33(11):1318--1324.

\bibitem[Horsburgh et~al., 2013]{horsburgh13}
Horsburgh, C.~R., Shea, K.~M., Phillips, P., and Lavalley, M. (2013).
\newblock {{R}andomized clinical trials to identify optimal antibiotic
  treatment duration}.
\newblock {\em Trials}, 14:88.

\bibitem[Iasonos et~al., 2008]{iasonos08}
Iasonos, A., Wilton, A.~S., Riedel, E.~R., Seshan, V.~E., and Spriggs, D.~R.
  (2008).
\newblock {{A} comprehensive comparison of the continual reassessment method to
  the standard 3 + 3 dose escalation scheme in {P}hase {I} dose-finding
  studies}.
\newblock {\em Clin Trials}, 5(5):465--477.

\bibitem[Le~Tourneau et~al., 2009]{letourneau09}
Le~Tourneau, C., Lee, J.~J., and Siu, L.~L. (2009).
\newblock Dose escalation methods in phase i cancer clinical trials.
\newblock {\em JNCI: Journal of the National Cancer Institute}, 101(10):708.

\bibitem[Nimmo et~al., 2015]{nimmo15}
Nimmo, C., Lipman, M., Phillips, P.~P., McHugh, T., Nunn, A., and Abubakar, I.
  (2015).
\newblock {{S}hortening treatment of tuberculosis: lessons from fluoroquinolone
  trials}.
\newblock {\em Lancet Infect Dis}, 15(2):141--143.

\bibitem[O'Quigley et~al., 1990]{oquigley90}
O'Quigley, J., Pepe, M., and Fisher, L. (1990).
\newblock Continual reassessment method: A practical design for phase 1
  clinical trials in cancer.
\newblock {\em Biometrics}, 46(1):33--48.

\bibitem[Phillips et~al., 2015]{phillips15}
Phillips, P. P.~J., Morris, T.~P., and Walker, A.~S. (2015).
\newblock Door/radar: A gateway into the unknown?
\newblock {\em Clinical Infectious Diseases}, 62(6):814.

\bibitem[Rehal et~al., 2016]{rehal16}
Rehal, S., Morris, T.~P., Fielding, K., Carpenter, J.~R., and Phillips, P.
  P.~J. (2016).
\newblock Non-inferiority trials: are they inferior? a systematic review of
  reporting in major medical journals.
\newblock {\em BMJ Open}, 6(10).

\bibitem[Royston and Altman, 1994]{royston94}
Royston, P. and Altman, D.~G. (1994).
\newblock Regression using fractional polynomials of continuous covariates:
  Parsimonious parametric modelling.
\newblock {\em Journal of the Royal Statistical Society. Series C (Applied
  Statistics)}, 43(3):429--467.

\bibitem[Royston et~al., 1999]{royston99}
Royston, P., Ambler, G., and Sauerbrei, W. (1999).
\newblock {{T}he use of fractional polynomials to model continuous risk
  variables in epidemiology}.
\newblock {\em Int J Epidemiol}, 28(5):964--974.

\bibitem[Snapinn, 2000]{snapinn00}
Snapinn, S.~M. (2000).
\newblock {{N}oninferiority trials}.
\newblock {\em Curr Control Trials Cardiovasc Med}, 1(1):19--21.

\bibitem[Sydes et~al., 2012]{sydes12}
Sydes, M.~R., Parmar, M.~K., Mason, M.~D., Clarke, N.~W., Amos, C., Anderson,
  J., de~Bono, J., Dearnaley, D.~P., Dwyer, J., Green, C., Jovic, G., Ritchie,
  A.~W., Russell, J.~M., Sanders, K., Thalmann, G., and James, N.~D. (2012).
\newblock {{F}lexible trial design in practice - stopping arms for
  lack-of-benefit and adding research arms mid-trial in
  {S}{T}{A}{M}{P}{E}{D}{E}: a multi-arm multi-stage randomized controlled
  trial}.
\newblock {\em Trials}, 13:168.

\bibitem[Ventola, 2015]{ventola15}
Ventola, C.~L. (2015).
\newblock {{T}he antibiotic resistance crisis: part 1: causes and threats}.
\newblock {\em P T}, 40(4):277--283.

\end{thebibliography}

\end{document}